\documentclass[aps,pre,preprint,showpacs,groupedaddress]{revtex4}
\usepackage{graphicx}

\newlength\figurewidth
\begin{document}

\title{Amplitude death of coupled hair bundles with stochastic channel noise}
\author{Kyung-Joong Kim and Kang-Hun Ahn}%
\affiliation{Department of Physics, Chungnam National University,
 Daejeon, 305-764, Republic of Korea}

\begin{abstract}
 Hair cells conduct auditory transduction in vertebrates.
In lower vertebrates such as frogs and turtles, due to the active
mechanism in hair cells, hair bundles(stereocilia) can be
spontaneously oscillating or quiescent. Recently, the amplitude
death phenomenon has been proposed [K.-H. Ahn, J. R. Soc. Interface,
{\bf 10}, 20130525 (2013)] as a mechanism for auditory transduction
in frog hair-cell bundles, where sudden cessation of the
oscillations arises due to the coupling between non-identical hair
bundles.
The gating of the ion channel is intrinsically stochastic due to the
stochastic nature of the configuration change of the channel. The
strength of the noise due to the channel gating
can be
 comparable to the thermal
Brownian noise of hair bundles.
 Thus, we perform stochastic simulations of the
elastically coupled hair bundles. In spite of stray noisy
fluctuations due to its stochastic dynamics, our simulation shows
the transition from collective oscillation to amplitude death as
inter-bundle coupling strength increases. In its stochastic
dynamics, the formation of the amplitude death state of coupled hair
bundles can be seen as a sudden suppression of the displacement
fluctuation of the hair bundles as the coupling strength increases.
The enhancement of the signal-to-noise ratio through the amplitude
death phenomenon is clearly seen in the stochastic dynamics. Our
numerical results demonstrate that the multiple number of
transduction channels per hair bundle is an important factor to the
amplitude death phenomenon, because the phenomenon may disappear for
a small number of transduction channels due to strong gating noise.


\end{abstract}
\maketitle



\section{INTRODUCTION}
The ear amplifies incoming sound signals with high sensitivity and
sharp frequency selectivity. It can detect sound stimuli over many
orders of magnitude in sound amplitude. The great sensitivity of
hearing originates from the active process in hair cells which
perform mechano-transduction in the inner ear. Sharp frequency
selectivity and active amplification arise with the aid of outer
hair cell(OHC) and membrane dynamics in the mammalian cochlea,
whereas non-mammals lacking in OHC such as frogs, also have acute
hearing relying on hair-cell bundle
dynamics\cite{fettiplace,maoileidigh,Nadrowski,Martin}. Hair bundles
can be spontaneously oscillating or quiescent depending on their
physical parameters. It has been speculated that hair bundles
operate on the border of these two regions, which is called
Hopf-bifurcation critical point\cite{equiluz,Nadrowski}. Such an
assumption may explain the high sensitivity and the sharp frequency
selectivity of hearing.

A question one may have is whether the bifurcation phenomenon indeed
exists in-vivo. In reality, it is more appropriate to assume a
distribution of physical parameters of hair bundles, rather than
assuming all hair bundles are equivalent as in a simple mathematical
model.  The dynamics of coupled hair bundles in a  homogeneous
configuration can be found in Ref.\cite{dierkes,dierkes2}. In a
recent theory, it has been demonstrated that parametric
distributions of hair bundles help auditory transduction through
inter-bundle coupling\cite{ahn}. The phenomenon has been known as
amplitude death in the nonlinear dynamics community, which has been
reported in many cases of coupled
oscillators\cite{Strogatz,ozden,Resmi,Pisarchik,Zhang,Ryu,Wei,Neufeld,Herrero,Hou,Karnatak,Prasad}.
The amplitude death phenomenon in hair cells has been introduced
through continuous variable dynamics\cite{ahn}, where the open
probabilities of ion channels are functions of the mechanical
displacement of the hair bundle.
Meanwhile, due to the stochastic nature of the configuration change
of the ion channel, the ion channel gating in cilia happens in a
stochastic manner with dwell time of 1 ms
\cite{Ricci2003,Clausznitzer}. This causes intrinsic noise
associated with the stochastic channel gating.

Furthermore, as we will show, the noise associated with stochastic
channel gating can be of comparable strength  to the thermal noise
force. Thus it is necessary to investigate whether the amplitude
death exists in the presence of the intrinsic channel gating noise.

In this work, we show that the amplitude death phenomenon of coupled
hair bundles indeed exists even when the bundle dynamics is treated
in a stochastic manner. We simulate the dynamics of the coupled hair
bundles by using the stochastic Markov process and find the features
of the amplitude death phenomenon for the coupled hair bundles,
where a rigorous model for bullfrog hair bundles has been used. We
find that the statistical distribution of the bundles' displacement
shows a transition from a bimodal to a single modal structure as we
increase the inter-bundle coupling strength. The displacement
distribution of hair bundles in the amplitude death state turns out
to be weakly asymmetric. The velocity distribution of the hair
bundles shows transition from the half-Lorentizian to Gaussian
distribution. In the region of spontaneous oscillation, the hair
bundle velocity distribution is asymmetric Lorentzian due to the
interplay between relaxation and the active process. In the
amplitude death region, velocity distribution is well described by
the Gaussian distribution. It is demonstrated that the
signal-to-noise ratio and the response of the coupled hair bundle to
an oscillatory stimulus is enhanced as the system is located in the
amplitude death region.

The relative importance of the stochastic channel gating noise is
discussed by comparing it to the thermal noise force on the hair
bundle displacement. We will show that the stochastic noise can be
reduced by increasing  the number of the ion channels per hair
bundle. For the hair bundles which have a few tens of ion channels,
we will show the stochastic channel gating noise is sufficiently
weakened so that  the amplitude death phenomenon survives both types
of noise.

 The paper is organised as follows.
 In Section II, we introduce the stochastic Markov model for the elastically coupled
 hair bundles. The calculation results for the stochastic hair bundle model are shown and discussed in Section III.
 We demonstrate the amplitude death in the stochastic dynamics in
 III A and the enhanced signal-to-noise ratio in III B.  The thermal
 noise and the stochastic channel-gating noise are compared and  the significance of the number of channels per hair bundle is discussed
 in III C.
 The conclusion is given in Section IV.

\section{STOCHASTIC DYNAMICS MODEL OF COUPLED HAIR BUNDLES}

Hair bundles in a bullfrog sacculus are coupled by the otolithic
membrane which has a finite mass\cite{Michael}. We consider a
one-dimensional hair bundle array which is elastically coupled with
finite mass elements attached to it in Fig. 1 (a). The equations
governing the dynamics of the coupled hair bundles are
\begin{eqnarray}
\label{main}
 m\ddot{x_{i}}=-m\gamma_{m}\dot{x_{i}} +
 k(x_{i+1}-2x_{i}+x_{i-1})+f_{\rm HB, \it i}
\end{eqnarray}
where $x_i$ is the displacement of $i$-th hair
bundle($i=1,2,3,\cdot\cdot\cdot,N$) and $N$ is the number of hair
bundles. Here, we use finite elements of mass $m$ to describe the
loading effect of the overlying membrane. $k$ is the inter-bundle
coupling strength and $\gamma_m$ is the friction constant per mass,
which was chosen to mimic the poor frequency selectivity of the
overlying membrane\cite{ahn}.

The force exerted on the $i$-th element by the $i$-th hair bundle is
\begin{eqnarray}
f_{\rm HB, \it i}=
-\lambda\dot{x_{i}}-k_{gs}(x_{i}-x_{a,i}-\frac{D}{N_{ch}}\sum_{j=1}^{N_{ch}}G_{i,j}
) -k_{sp,i}x_{i},
\end{eqnarray}
where the pivotal stiffness and the combined gating spring stiffness
are denoted by $k_{sp}$, and $k_{gs}$, respectively. $D$ is the
displacement associated with channel gating and $\lambda$ is the
friction coefficient of the hair bundle\cite{Nadrowski}. To simulate
the dynamics of the multiple number of transduction channels per
hair bundle, we assume $N_{\rm ch}$ number of transduction channels.
 For simplicity, we assume that all the molecular motor
positions of the $i$-th hair bundle are equal to $x_{a,i}$. Here,
$G_{i,j}=1(0)$ when the $j$-th channel of $i$-th bundle is open
(closed). The probability for $G_{i,j}$ to be 1 is given by the open
probability of the $i$-th ion channel, $p_{o,i}$. $x_{a,i}$ is the
position of the molecular motor in the $i$-th hair bundle, which
satisfies
\begin{eqnarray}
\label{xa}
\lambda_{a}\dot{x}_{a,i}=k_{gs}(x_{i}-x_{a,i}-\frac{D}{N_{ch}}\sum_{j=1}^{N_{ch}}
G_{i,j})-f_{max,i}(1-Sp_{o,i}),
\end{eqnarray}
where $f_{max,i}(1-Sp_{o,i})$ is the force exerted by the molecular
motor and $\lambda_{a}$ is the velocity-force relation constant of
the molecular motor\cite{Nadrowski}. $f_{max,i}$ is the motor's
maximal force and $S$ is a parameter for the strength of the calcium
feedback\cite{Nadrowski}. Since the feedback force strength is a
function of the calcium ion concentration in the stereocilia, we
think it is more natural to assume that the feedback force is a
function of the mean open probability $p_{o,i}$ rather than the
number of open channels at a particular time.

In vivo, the open-closed transitions of the ion channels in cilia
occurs stochastically with a dwell time of 1 ms\cite{Ricci2003},
which is our motivation to describe our model for coupled hair
bundles in the Markov process.

See Fig. 1. Let us consider a two-state channel which has the energy
$E_o$ in the open state and $E_c$ in the closed state. The
equilibrium transition rates of these two states, $\Gamma_{oc}$ and
$\Gamma_{co}$, are related by the Boltzmann equation
$\frac{\Gamma_{oc}}{\Gamma_{co}}=e^{-(E_o-E_c)/k_{B}T}=e^{-\Delta
E/k_{B}T}$ ( $k_B$ is the Boltzmann constant, and $T$ is the
temperature) where $\Delta E$ is the energy difference associated
with the channel gating. The evolution of a continuous-time process
is given by the first-order differential equation
\begin{eqnarray}
\frac{dp_o}{dt}=\Gamma_{oc}(1-p_{o})-\Gamma_{co}p_o.
\end{eqnarray}
In the case of $\frac{dp_o}{dt}\cong0$, we obtain the open
probability by
\begin{eqnarray}
p_o\cong\frac{\Gamma_{oc}}{\Gamma_{co}+\Gamma_{oc}}=1/(1+e^{\Delta
E/k_{B}T}).
\end{eqnarray}
 The channel gating in hair cells is caused by
mechanical stimulation which is delivered by a tip link, a filament
interconnecting adjacent stereocilia\cite{peter}. Since the
mechanical energies stored in the gating springs are
$E_o=\frac{1}{2}k_{gs}(x-x_a-D)^2$ and
$E_c=\frac{1}{2}k_{gs}(x-x_a)^2$, the open probability of the $i$-th
hair bundle is given by
\begin{eqnarray}
p_{o,i}&=&1/(1+A\exp(-\frac{x_{i}-x_{a,i}}{\delta}))
\end{eqnarray}
where $A$ is a constant associated with the intrinsic free energy
difference between the open and closed states and
$\delta=k_{B}TN_{ch}/(k_{gs}D)$ is a typical length associated with
the channel gating\cite{Nadrowski,Hudspeth,Mehta}. Note that this
length $\delta$ is independent of the number of channel $N_{ch}$,
because the combined gating spring stiffness $k_{gs}$ is
proportional to $N_{ch}$.

At each time step with interval $\Delta t$, we generate  random
numbers $\xi_{i,j}$ ($i=1,2,\cdots, N$ and $j=1,2,\cdots, N_{ch}$)
which are distributed between 0 and 1, and we compare then with the
transition probabilities,
 \begin{eqnarray}
\omega_{oc,i} &=& \gamma\Delta t p_{o,i} \\
\omega_{co,i} &=& \gamma\Delta t (1-p_{o,i})
\end{eqnarray}
where $\gamma$ is a parameter giving the channel relaxation rate.
When the channel is in the open state ($G_{i,j}=1$), it remains in
the open state  if $\omega_{co,i} < \xi_{i,j}$, or changes to the
closed state ($G_{i,j}=1\rightarrow G_{i,j}=0$) if $\omega_{co,i} >
\xi_{i,j}$. Similarly, when the channel is in the closed state, the
state changes to the open state($G_{i,j}=0$), if $\omega_{oc,i} >
\xi_{i,j}$ ($G_{i,j}=0\rightarrow G_{i,j}=1$).

The thermal fluctuation of the channel gate affects the bandwidth of
the intrinsic stochastic noise. The fluctuation-dissipation theorem
says that as the fluctuation gets larger, it accompanies a larger
dissipation. Since, in our simulation, the bandwidth of the channel
gating noise is given by $1/\Delta t$, and the relaxation rate of
the gating is about $\gamma$, these two values should be in the same
order of magnitude to fulfill the fluctuation dissipation theorem.
Thus, throughout this work, we choose $\gamma \Delta t=0.4$. As will
be shown here, our stochastic simulation results resemble those of
the continuous variable calculations\cite{ahn} for a sufficiently
large relaxation rate $\gamma$ or a large number of ion channels per
bundle $N_{ch}$, which is likely the case in biological systems. In
contrast, when the relaxation of the ion channel is not fast enough
the channel gating noise becomes too strong and might destroy the
amplitude death phenomenon.

\section{RESULTS AND DISCUSSION}
As discussed in Ref.[1], an inhomogeneous distribution of the
physical parameters of hair bundles is likely to exist in biological
systems and is critical to the emergence of amplitude death. We
simulate the non-uniformity of the hair bundles by using a
distribution of the bundle's pivotal stiffness $k_{sp}$ and the
maximal motor force $f_{max}$ (Fig. 2 (a)). The parameters for hair
bundles are given to ensure that some of them are closed and the
others are spontaneously oscillating.

\subsection{SUPPRESSION OF MECHANICAL FLUCTUATION OF COUPLED HAIR BUNDLES}

To investigate the existence of amplitude death in a stochastic
nature, it proves useful to calculate the histogram for the
displacement of the hair bundles as shown in Fig. 2 (b)$\sim$(g).
These histograms depict the number of events for the hair bundles to
be located in the given displacement interval. The events are
counted every millisecond for 2 seconds. In the absence of
coupling(Fig. 2 (b)), the histogram shows a multiple-peak structure,
which represents the different oscillation states or closed states
for each of the hair bundles. Since we assume that the ion channels
are either closed or oscillatory, the peaks are mainly on the
position of the closed state rather than on the open state. At weak
couplings(Fig. 2 (c),(d),(e)), the histograms show two main peaks.
As the coupling strength increases, the distance between these two
main peaks becomes reduced. Above a certain coupling strength, the
histogram shows only one peak corresponding to the closed state(Fig.
2 (f),(g)). The tail of the peak in the histogram rapidly disappears
as the strength increases further(Fig. 2 (g)). This corresponds to
the amplitude death phenomenon of the coupled hair bundles which has
been shown in the continuum model for hair bundles\cite{ahn}.

The transition to the amplitude death region can also be seen
through the velocity distribution.
 We find that it
has a half-Lorentzian distribution in the collectively oscillating
states and shows a transition to Gaussian distribution(Fig. 3). The
underlying mechanism for the half-Lorentzian distribution arises
from the bundles opening slowly and closing quickly due to the
relaxation process and active oscillation(inset of Fig. 3). In the
amplitude death region, random channel gating noise causes the
Gaussian distribution of the velocity.

 In Fig. 4, we show the variances for the displacements of the
 coupled hair bundles which are calculated from the stochastic
 and continuum model. The variance $\sigma_{X}^{2}$
  is the average of the
variance of the displacement of each single hair bundle
\begin{eqnarray}
\sigma_{ X}^2&=&\frac{1}{N}\sum_{i}\sigma^{2}_{{ X},i} \\
\sigma^2_{X,i}&=&<(x_i-<x_i>_t)^2>_{t},
\end{eqnarray}
where $< >_{t}$ means the average over time.
  The error bars in Fig. 4
 are estimated from 20 trials of the simulations. The $\sigma_{X}$
value shows a good agreement with the values which were obtained
from a continuum model when the coupling is not sufficiently strong
enough to cause amplitude death. When the coupling strength is
sufficiently strong, the fluctuation of coupled hair bundles in the
stochastic model also shows a dramatic reduction as in the continuum
model calculation of Ref.[1]. Thus, even in the presence of the
channel gating fluctuations, the strong reduction of the mechanical
fluctuation appears as the fingerprint of the amplitude death
phenomenon, which was originally considered in noise-free
dynamics\cite{Strogatz,ozden,Resmi,Pisarchik,Zhang,Ryu,Wei,Neufeld,Herrero,Hou,Karnatak,Prasad}.

While the transition to the amplitude death region is clearly seen
in the distribution of the displacement and velocity, the normalized
correlation between the hair bundles does not have any significant
difference in these two regions.
 We define the correlation between hair bundle
displacements  by  $ C=\frac{2}{N}\sum_{i\neq
j}\{<x_ix_j>_t-<x_i>_t<x_j>_t\}. $ We find that this correlation
function is also suppressed in the amplitude death region (see
Supplemental Material Fig. 9 (a)). We define the normalized
correlation function $C_N$ for the displacement of the coupled hair
bundles which is given by $ C_N\equiv [1+2(\frac{\sum_{i\neq
j}(<x_ix_j>_t-<x_i>_t<x_j>_t)}{\sum_i(<x_i^2>_t-<x_i>_t^2)})]/N.$
Then,
 $C_N$ is 0.1 at zero coupling strength but it goes up to 0.8 as $k$
increases, and retains the value regardless of amplitude death (see
Supplemental Material Fig. 9 (b)). This means that movements of the
hair bundles are rather coherent and they do not lose this coherence
in the amplitude death region. In other words, the amplitudes of
oscillations are quenched but their cross-correlation is maintained
in the amplitude death region.

\subsection{ ENHANCEMENT OF SIGNAL-TO-NOISE RATIO BY AMPLITUDE DEATH}

Now we show how the hair bundles in the amplitude death state
respond to oscillatory stimuli.
 We investigate the time evolution of the average
displacement (Fig. 5 (a))  of the coupled hair bundles in our
stochastic model. We applied a periodic stimulus $F_{ext}$= 2 pN
$\times \sin(2\pi\times 10 {\rm Hz}\times t)$ only when 2 sec $< t <
$ 4 sec.
 In the absence of any stimulus, we can see the coupled hair bundles move collectively
 for the coupling  $k = 2$ pN/nm. Then the collective oscillation is
destroyed above a certain coupling strength($k = 4.3$ pN/nm) where
the critical coupling strength for the transition to amplitude death
is not universal but depends on the hair bundles' parameter
distribution (not shown). Fig. 5 (b) shows the time evolution of the
ratio of open to total channels, which is defined by
\begin{eqnarray}
\tilde{G_{o}}&=&\frac{1}{N}\sum^{N}_{i=1}G_{i},\\
G_{i}&=&\frac{1}{N_{ch}}\sum^{N_{ch}}_{j=1}G_{i,j}.
\end{eqnarray}
This shows a similar pattern with the averaged open probability
$\tilde{P_{o}}=\frac{1}{N}\sum_{i}p_{o,i}$ in the continuum
dynamics.

 It is interesting to
note that the fluctuation of $X$ (Fig. 5 (a)) and $\tilde{G_o}$
(Fig. 5 (b)) when $k=6$ pN/nm is lower than the value for the
non-coupling case $k=0$.
 They  fluctuate much less if the system is in the amplitude death state
($k = 6$ pN/nm) compared to the non-coupling ($k = 0$) or
collectively oscillating state($k = 2$ pN/nm). With the periodic
stimulus $F_{ext} \neq 0$, the response of the amplitude death state
($k$ = 6 pN/nm) is much stronger than the uncoupled case ($k$ = 0),
while it is slightly weaker than the collectively oscillating case
($k$ = 2 pN/nm).

The signal amplification and noise reduction which is shown in Fig.
5 can be quantified by the signal-to-noise ratio (SNR).
 We calculate the power spectra of the mechanical displacement of the
hair bundles,
\begin{eqnarray}
S_{X}(f)=\frac{1}{N}\sum_{i}^{N}|X_{i}(f)|^2,
|X_{i}(f)|=\frac{1}{T_a}\int^{T_a}_{0}x_{i}(t)e^{i 2\pi  f t}dt
\end{eqnarray}
where $T_a$ is the time period for the Fourier transformation.
\textit{i} is the index of a hair bundle. We plot $S_{X}(f)$ for a
pure tone signal ($f = 12$ Hz) of amplitude 0.2 pN. When a weak
signal is applied to the uncoupled and weakly coupled bundles, the
signal is buried in the noisy fluctuations as shown in Fig. 6
((a),(b)). But, if the acoustic signal is applied to
strongly-coupled hair bundles(Fig. 6 (c)), the signal gets clearly
exposed due to the reduction of the noisy fluctuations. SNR is
defined by
\begin{eqnarray}
SNR = \lim_{\Delta f \rightarrow 0}\frac{S_{X}(f)}{\frac{1}{\Delta
f}\int_{f-\Delta f/2}^{f+\Delta f/2}S_{X}(f^{\prime})df^{\prime}},
\end{eqnarray}
where $f$ is the frequency of the external stimulus.
$F_{ext}(t)=F\sin(2 \pi f t)$ is added on the right hand side of
Eq.(1). We obtained the power spectra averaged over 50 different
trials.

In Fig. 6 (d), we show that the SNR tends to increase with the
coupling strength $k$. Due to the amplitude death, the increase of
the SNR is enhanced. For the case of $N_{ch}=20$, $\gamma=10$ ${\rm
ms}^{-1}$, which has been used so far, one can see this enhancement
of the SNR as the coupling strength $k$ increases. The increases of
SNR is more enhanced as $k$ goes beyond 4 pN/nm, which enters the
amplitude death region.

Fig. 6 (d) (and Supplemental Material Fig. 7) shows that the channel
gating noise becomes weaker as the channel relaxation rate $\gamma$
increases and the number of channels per bundle $N_{ch}$ decreases.
Thus, the SNR for $\gamma$ = 200 ms$^{-1}$ with single channel
$N_{ch}=1$ looks similar for the bundles with 20 channels with
$\gamma$ = 10 ms$^{-1}$. Thus, if the hair bundle has only one
channel with the usual relaxation rate $\gamma$ = 10 ms$^{-1}$, then
we do not see the amplitude death phenomenon due to the strong
noise. In this case, the enhancement of the SNR is not significant
as shown in Fig. 6 (d).

\subsection{ THERMAL NOISE vs TRANSDUCTION CHANNEL GATING NOISE}

The  sources of the noise can be divided into those which arise from
thermal fluctuation associated with the Brownian motion of hair
bundles, and fluctuation associated with the stochastic nature of
channel gating. Fluctuation associated with
the stochastic activity of molecular motors
 also exists but its strength is relatively
weak compared to the thermal noise\cite{Nadrowski}, which we do not
discuss here. While the thermal noise was considered in an earlier
work\cite{ahn}, we have investigated the role of the stochastic
channel noise so far in this work.  Now, let us compare the two
different noise sources, by including the thermal noise
force\cite{ahn}. This can be done by replacing Eq.(\ref{main}) with
\begin{eqnarray}
\nonumber
 m\ddot{x_{i}}&=&-m\gamma_{m}\dot{x_{i}} +
 k(x_{i+1}-2x_{i}+x_{i-1})+f_{\rm HB, \it i}\\ &+&f_{N,i}(t),
\end{eqnarray}
$f_{N,i}(t)$ is the thermal noise force which  is exerted on the
$i$-th hair bundle and satisfies the equipartition theorem. The
noise force has the finite correlation time $\tau_{c}$ and it has
the correlation function in the form of
\begin{eqnarray}
<\overline{f_{N,i}(t)f_{N,j}(t+\tau)}>=
\frac{2k_BT\tilde{\lambda}\delta_{i,j}}{e^{(\frac{\tilde{\lambda}\tau_c}{2m})^2}\textrm{erfc}(\frac{\tilde{\lambda}\tau_c}{2m})}
\frac{1}{\sqrt{\pi}\tau_c}e^{-(\tau/\tau_c)^2},
\end{eqnarray}
where $\tilde{\lambda}=\lambda+m\gamma_{m}$ and erfc(x) is the
complementary error function. (see Supplemental Material of
Ref.\cite{ahn}) $\tau_c$ is the correlation time of the thermal
noise force which acts on the mechanical motion of the hair bundle.
This value in many cases is assumed to be zero for the simplicity of
the calculation. In reality, $\tau_c$ depends on the detail of the
environment of the hair bundle. In this work, we choose a value
close to a channel relaxation rate $\gamma^{-1}$.

Fig. 7 (a) shows that the channel gating noise can be more or less
important to the mechanical fluctuations $\sigma_{X}$ compared to
the thermal noise depending on the number of channels per bundle.
Our numerical estimation for $N_{ch}=20$ shows that the mechanical
noise power $S_{X}(f)$ due to the channel gating is weaker than the
noise due to the thermal noise force (see Supplemental Material Fig.
4, 5, and 6). However, this situation is opposite when we consider
the fluctuation of the number of open channels $\sigma_{G}$, which
is important to the transduction ion current in a neuronal signal.
$\sigma_{G}$ is defined by
\begin{eqnarray}
\sigma_{ G}^2&=&\frac{1}{N}\sum_{i}\sigma^{2}_{{ G},i} \\
\sigma^2_{G,i}&=&<(G_i-<G_i>_t)^2>_{t}.
\end{eqnarray}

We find that the channel gating noise is the more important noise
source to the number of open channels compared to the thermal
Brownian noise force (Fig. 7 (b)). The channel gating noise
contribution to $\sigma_{G}$ becomes weakened as $N_{ch}$ increases.
Thus, for the case of $N_{ch}=20$, we see that the thermal noise
force and the channel gating noise  are in comparable strength. When
the two noise sources coexist, the fluctuation of hair bundles is
given by the sum of the contribution of each source (see
Supplemental Material Fig. 7).


\section{CONCLUSION}

 We developed a stochastic model for the dynamics
of the coupled hair bundles with channel gating noise. As in the
continuum model for hair bundle dynamics[1], there exists a
transition from the collective oscillation state to the amplitude
death state above a certain coupling strength. The transition is
also evidenced by the fact that the velocity distribution change
from the half-Lorentzian distribution to the Gaussian distribution.
The relative importance of the stochastic channel gating noise is
compared to the thermal noise force on the hair bundle displacement
and the number of open channels. We find that the stochastic channel
noise can be reduced by increasing the number of the ion channels
per hair bundle. For the hair bundles which have a few tens of the
ion channels, we conclude that the stochastic channel gating noise
is sufficiently weak so that the amplitude death phenomenon
survives. Therefore, the enhancement of the signal-to-noise ratio of
coupled hair bundles would be possible through the amplitude death
phenomenon even with the stochastic channel noise.

\begin{acknowledgements}
This work was supported by the National Research Foundation of Korea
(NRF) grant funded by the Korea government(MEST) (No.2011-0009557).
We thank Jung-Wan Ryu, Jan Wiersig, and Joel Rasmussen for helpful
discussion and careful reading of our manuscript.
\end{acknowledgements}


\newpage
\begin{figure} \includegraphics[width=0.5\textwidth]{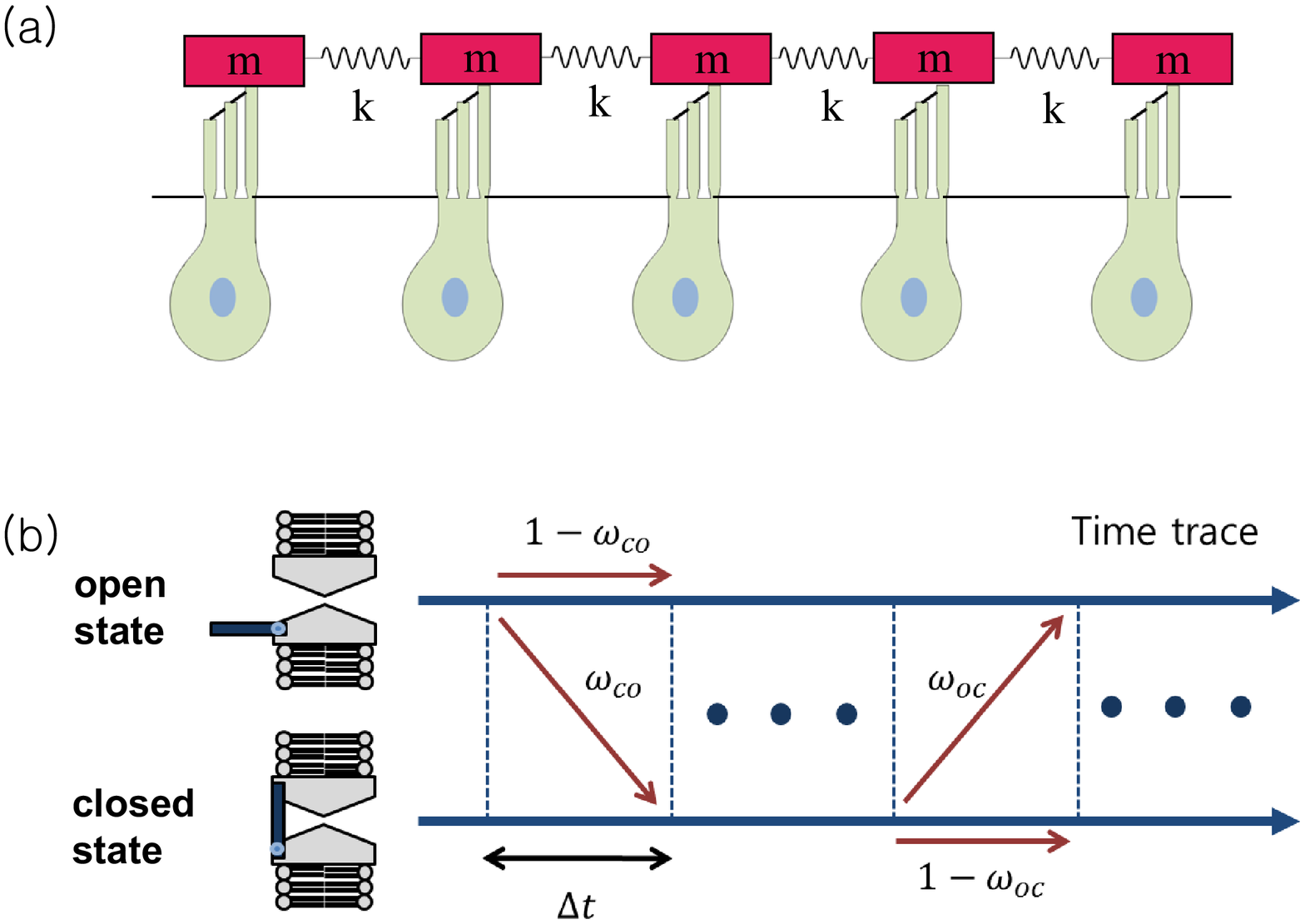}
\caption{(a) A model for coupled hair bundles. (b) Schematic drawing
for Markov analysis of stochastic simulation for the coupled hair
bundles dynamics.} \label{Fig1}
\end{figure}

\newpage
\begin{figure} \includegraphics[width=0.5\textwidth]{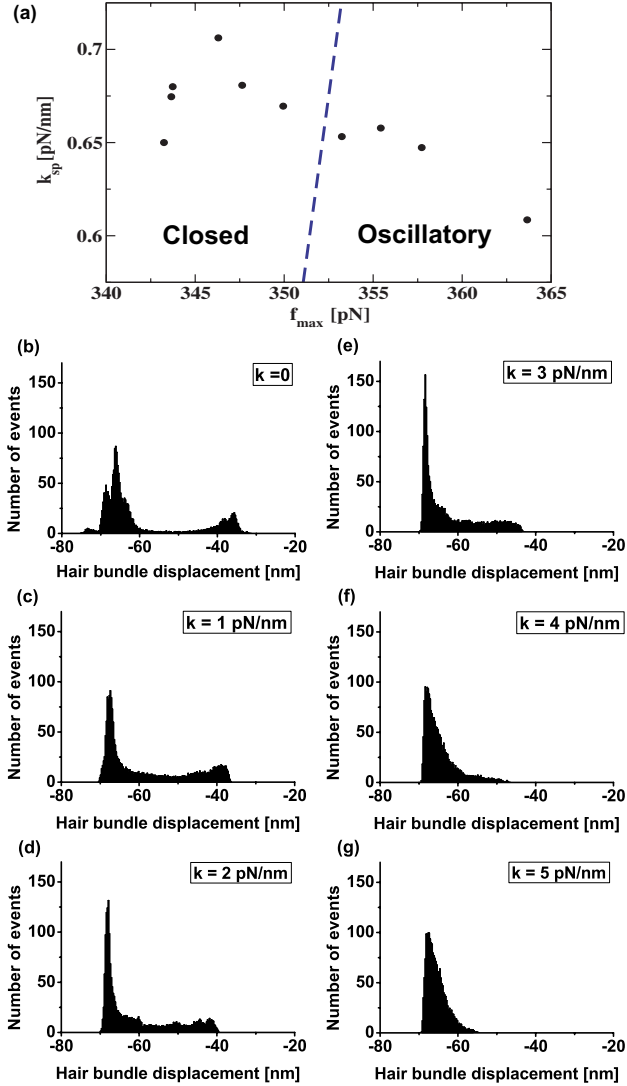}
\caption{(a) The parameter distribution for the pivotal stiffness of
hair bundles $k_{sp}$ and the maximum forces $f_{max}$ of molecular
motors. We use this distribution for all simulations in this paper.
(b)$\sim$(g) The histograms show the number of events which are
counted every 1 millisecond for 2 seconds for (b) incoherent ($k$ =
0), (c)$\sim$(f) locking (or collectively oscillating) ($k$ = 1
pN/nm, $k$ = 2 pN/nm, $k$ = 3 pN/nm, and $k$ = 4 pN/nm), and (g)
amplitude death region ($k$ = 5 pN/nm) of 10 hair bundles.}
\label{Fig2}
\end{figure}

\newpage
\begin{figure} \includegraphics[width=0.41 \textwidth]{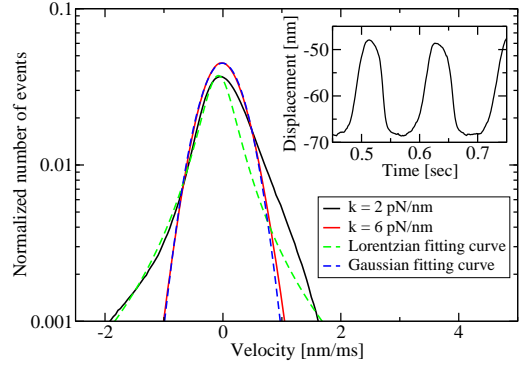}
\caption{ The velocity distribution for amplitude death($k$ = 6
pN/nm), locking($k$ = 2 pN/nm) of 10 coupled hair bundles. The green
dashed line is a fitting curve for the velocity distribution for
locking oscillation(the black solid line) which is a half-Lorentzian
curve. The blue dashed line is a Gaussian fitting curve for the
velocity distribution for the amplitude death state (the red solid
line). [inset] The averaged displacement of the coupled hair bundles
at $k$ = 2 pN/nm as a function of time.} \label{Fig3}
\end{figure}

\newpage
\begin{figure} \includegraphics[width=0.5\textwidth]{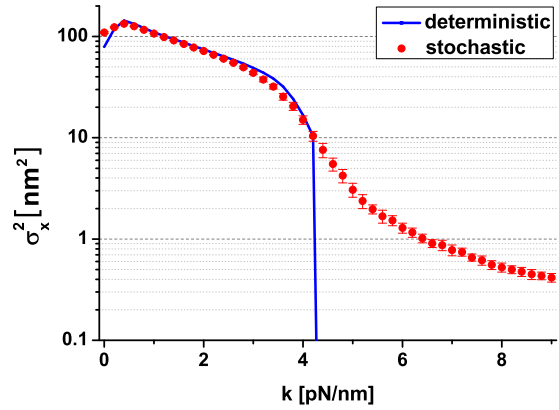}
\caption{ The averaged variance for the displacement of the coupled
hair bundles as a function of the inter-band coupling strength $k$.}
\label{Fig4}
\end{figure}

\newpage
\begin{figure} \includegraphics[width=0.5\textwidth]{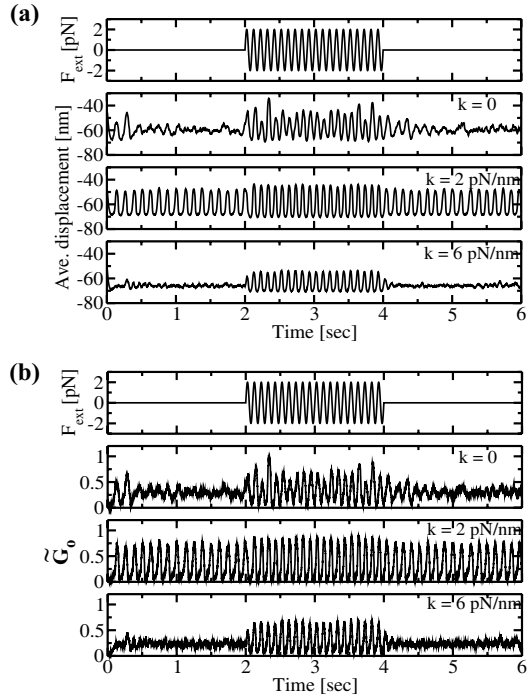}
\caption{(a) The average displacements $X=\frac{1}{N}\sum_{i}X_{i}$
for amplitude death($k$ = 6 pN/nm), collectively oscillating ($k$ =
2 pN/nm), and uncoupled ($k$ = 0) states of 10 coupled hair bundles.
The top of the figure is the external force acting on the hair
bundles. (b) The open channel ratio ${\tilde G}_{o}$ of 10 coupled
hair bundles. } \label{Fig5}
\end{figure}

\newpage
\begin{figure} \includegraphics[width=0.5\textwidth]{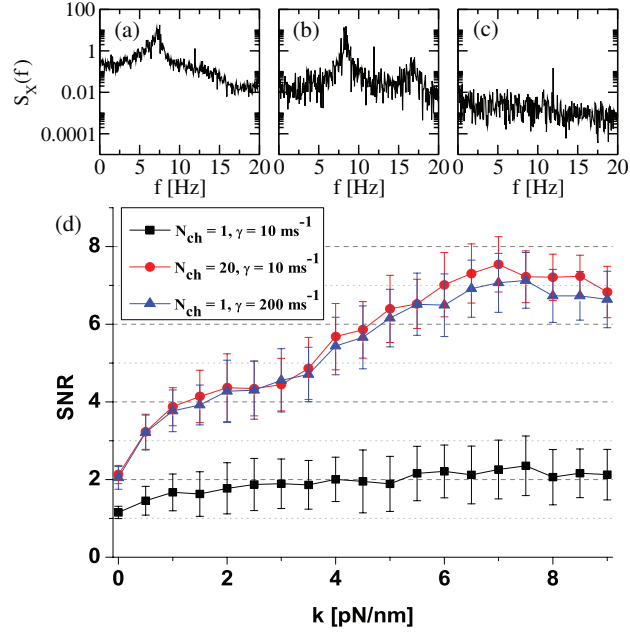}\\
\caption{(a-c) The power spectra of mechanical displacement $S_X(f)$
as a function of frequency $f$ when an external force is applied.
The coupling strengths are (a)$k = 0$, (b)$k = 3$ pN/nm, (c)$k = 9$
pN/nm. (d) SNR as a function of the coupling strength $k$. We used a
0.2 pN external force with frequency of 12 Hz in our simulations
(a-e). We obtained the power spectra averaged over 50 different
trials. } \label{Fig6}
\end{figure}

\newpage
\begin{figure} \includegraphics[width=0.4\textwidth]{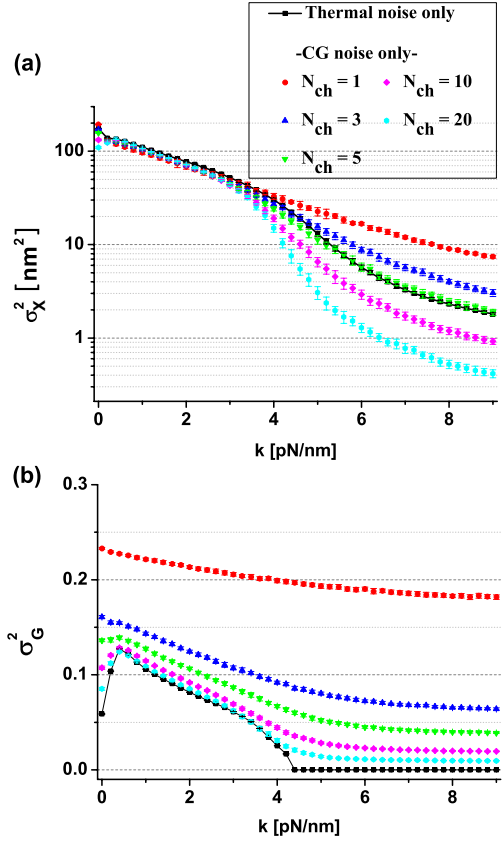}
\caption{ (a) The variance $\sigma_{x}$ of the hair bundles'
displacement as a function of the coupling strength $k$ (b) The
variance $\sigma_{G}$ of the number of the open transduction
channels as a function of the coupling strength.  The black square
dots are for the case when only the thermal noise exists. When only
channel gating noise exists we plot the variances for the various
number of channels. The error bars are obtained by 50 different
trials. We used a correlation time $\tau_c =$ 0.14 ms for the
thermal noise and a relaxation
rate $\gamma =$ 10 ms$^{-1}$ for the stochastic channel gating
noise. In both case, as the number of ion channels per hair bundle
increases, the variance decreases.} \label{Fig7}
\end{figure}

\newpage

\begin{table}[p]
  \centering
\begin{tabular}{ p{2cm} p{9cm} p{3.5cm} l }
\hline
  parameter & definition & value & Ref. \\
 \hline
  $m$ &  mass of one unit of cross-cut membrane & 2 $\mu$g &
   \\
  $\gamma_m$ & friction constant per mass of the membrane & 0.5 ms$^{-1}$ &  \\
  $k$ & inter-bundle coupling strength & 0 $-$ 9 pN nm$^{-1}$ &  \\
  $N$ & size of the one-dimensional hair bundle array & 10 &  \\
  $N_{ch}$ & The number of channels per hair bundle & 20 &  \\
  $\lambda$ & friction of a hair bundle & 2.8 $\mu$N s m$^{-1}$ & \cite{Nadrowski} \\
  $\lambda_a$ & friction of adaptation motors & 10 $\mu$N s m$^{-1}$ & \cite{Hacohen} \\
  $k_{gs}$ & combined gating spring stiffness & 0.75 pN nm$^{-1}$ & \cite{Mehta} \\
  $<k_{sp}>$ & mean value of hair bundle pivot stiffness & 0.65 pN nm$^{-1}$ & \cite{Mehta} \\
  $\delta k_{sp}$ & variance of hair bundle pivot stiffness & 0.05 pN nm$^{-1}$ &  \\
  $<f_{max}>$ & mean value of maximal motor force & 350 pN & \cite{Nadrowski} \\
  $\delta f_{max}$ & variance of maximal motor force & 7.14 pN &  \\
  $D$ & gating spring elongation & 60.9 nm & \cite{Mehta} \\
  $S$ &  strength of the calcium feedback & 0.65 & \cite{Nadrowski} \\
  $T$ & temperature & 300 K &  \\
  $A$ &  constant associated with the intrinsic free energy
difference & exp(16.7) & \cite{Nadrowski,Hudspeth} \\
  $\delta$ & typical
length associated with the channel gating & 4.53 nm & \cite{Nadrowski,Hudspeth} \\
  $\gamma$ &  channel relaxation rate & 10 ms$^{-1}$ &  \\
  $\Delta t$ &  time interval for stochastic simulation  & 0.04 ms &  \\
  \hline
\end{tabular}
 \caption{List of the parameter values for the simulations}\label{table1}
\end{table}

\end{document}